\documentclass[fleqn,usenatbib]{mnras}

\usepackage{newtxtext,newtxmath}

\usepackage[T1]{fontenc}
\usepackage{ae,aecompl}
\usepackage{cleveref}

\usepackage{graphicx}	
\usepackage{amsmath}	
\usepackage{amssymb}	

\newcommand\target{{MAXI~J1348$-$630} }
\newcommand\nicer{{\it NICER} }
\newcommand\maxi{{\it MAXI} }

\usepackage{xcolor}

\title[A NICER view of MAXI~J1348$-$630]{NICER observations reveal that the X-ray transient MAXI~J1348$-$630 is a Black Hole X-ray binary}

\author[Zhang et al.]{L. Zhang$^{1},$\thanks{E-mail: liang.zhang@soton.ac.uk}
D. Altamirano$^{1}$,
V. A. C\'uneo$^{2,3}$,
K. Alabarta$^{1,4}$,
T. Enoto$^{5}$,
\newauthor
J. Homan$^{6,7}$,
R. A. Remillard$^{8}$,
P. Uttley$^{9}$,
F. M. Vincentelli$^{1}$,
Z. Arzoumanian$^{10}$,
\newauthor
P. Bult$^{10}$,
K. C. Gendreau$^{10}$,
C. Markwardt$^{10}$,
A. Sanna$^{11}$,
T. E. Strohmayer$^{10}$,
\newauthor
J. F. Steiner$^{12}$,
A. Basak$^{9}$,
J. Neilsen$^{13}$ \&
F. Tombesi$^{10,14,15,16}$
\\
$^{1}$Physics and Astronomy, University of Southampton, Southampton, Hampshire SO17 1BJ, UK \\
$^{2}$Instituto de Astrof\'isica de Canarias (IAC), V\'ia L\'actea s/n, La Laguna 38205, S/C de Tenerife, Spain \\
$^{3}$Departamento de Astrof\'isica, Universidad de La Laguna, La Laguna, E-38205, S/C de Tenerife, Spain \\
$^{4}$Kapteyn Astronomical Institute, University of Groningen, PO Box 800, NL-9700 AV Groningen, the Netherlands\\
$^{5}$Extreme Natural Phenomena RIKEN Hakubi Research Team, RIKEN Cluster for Pioneering Research, 2-1 Hirosawa, \\
Wako, Saitama 351-0198, Japan \\
$^{6}$Eureka Scientific, Inc., 2452 Delmer Street, Oakland, CA 94602, USA \\
$^{7}$SRON, Netherlands Institute for Space Research, Sorbonnelaan 2, 3584 CA Utrecht, The Netherlands \\
$^{8}$MIT Kavli Institute for Astrophysics and Space Research, MIT, 70 Vassar Street, Cambridge, MA 02139, USA \\
$^{9}$Anton Pannekoek Institute for Astronomy, University of Amsterdam, Science Park 904, 1098 XH Amsterdam, The Netherlands\\
$^{10}$Astrophysics Science Division, NASA Goddard Space Flight Center, Greenbelt, MD 20771, USA \\
$^{11}$Dipartimento di Fisica, Universit\`a degli Studi di Cagliari, SP Monserrato-Sestu km 0.7, 09042 Monserrato, Italy \\
$^{12}$Harvard-Smithsonian Center for Astrophysics, 60 Garden St., Cambridge, MA 02138, USA \\
$^{13}$Villanova University Department of Physics, Villanova, PA, 19085, USA \\
$^{14}$Department of Physics, University of Rome `Tor Vergata', Via della Ricerca Scientifica 1, I-00133 Rome, Italy \\
$^{15}$Department of Astronomy, University of Maryland, College Park, MD 20742, USA \\
$^{16}$INAF Osservatorio Astronomico di Roma, Via Frascati 33, 00078 Monte Porzio Catone (Rome), Italy 
}

\date{Accepted 2020 September 16. Received 2020 September 16; in original form 2020 April 09}

\pubyear{2019}

\begin{document}
\label{firstpage}
\pagerange{\pageref{firstpage}--\pageref{lastpage}}
\maketitle

\begin{abstract}
We studied the outburst evolution and timing properties of the recently discovered X-ray transient \target as observed with {\it NICER}. We produced the fundamental diagrams commonly used to trace the spectral evolution, and power density spectra to study the fast X-ray variability. The main outburst evolution of \target is similar to that commonly observed in black hole transients. The source evolved from the hard state, through hard- and soft-intermediate states, into the soft state in the outburst rise, and back to the hard state in reverse during the outburst decay. At the end of the outburst, \target underwent two reflares with peak fluxes $\sim1$ and $\sim2$ orders of magnitude fainter than the main outburst, respectively. During the reflares, the source remained in the hard state only, without undergoing any state transitions, which is similar to the  
so-called ``failed outbursts''. Different types of quasi-periodic oscillations (QPOs) are observed at different phases of the outburst. Based on our spectral-timing results, we conclude that \target is a black hole candidate.  
\end{abstract}




\section{Introduction}
\label{sec:intro}

Black hole transients (BHTs) usually show distinct spectral and temporal states during an outburst (see \citealt{Remillard2006} and \citealt{Belloni2011} for reviews). Following the classification of \citet{Homan2005}, four main states can be identified based on their spectral-timing properties: the hard state (HS), the  hard-intermediate state (HIMS), the soft-intermediate state (SIMS), and the soft state (SS). The outburst evolution along these states can be well studied through several fundamental tools, e.g., the hardness-intensity diagram (HID, \citealt{Homan2001}), the hardness-rms diagram (HRD, \citealt{Belloni2005}), and the rms-intensity diagram (RID, \citealt{Munoz2011}). During most outbursts, BHTs evolve from the HS, through the HIMS and SIMS, into the SS in the outburst rise. In the outburst decay, the outbursts generally evolve back to the HS in reverse order  \citep[see, e.g.,][]{Homan2005,Klis2006,Belloni2011,Motta2011}. Note that the transition between the different states does not necessarily follow the given order, as there can be excursions to either harder or softer states \citep{Belloni2011}.
In addition, some outbursts of some sources never undergo a transition to the SS. These outbursts are commonly known as ``failed outbursts" \citep[e.g.,][]{Sturner2005,Capitanio2009,Del2016}.   

Timing properties also change dramatically between states \citep[e.g.,][and references therein]{Motta2016}. In the HS and HIMS, the power density spectra (PDS) are dominated by strong band-limited noise and sometimes a type-C QPO. Type-C QPOs are characterized by a narrow peak with variable frequency ranging from several mHz to $\sim30$ Hz. A second harmonic and a subharmonic are usually observed. 
The SIMS is characterized by showing the so-called type-B QPOs: a QPO (generally with frequencies of 4--6 Hz), accompanied by weak (few percent fractional rms in the full {\it RXTE}/PCA band) broadband noise \citep[e.g.,][]{Casella2004,Motta2011,Motta2016,Stevens2016,Gao2017,Stevens2018}. 
The typical PDS of the SS does not reveal any strong variability (fractional rms of $\lesssim2\%$ in the full {\it RXTE}/PCA band). 
Less commonly but sometimes seen in the SS are the so-called type-A QPOs. These QPOs appear in a similar frequency range (6--8 Hz) as the type-B, but are broader \citep[e.g.,][]{Casella2004,Rodriguez2004,Motta2011,Sriram2013,Motta2016}.
And weak type-C QPOs are sometimes observed in the SS, where they reach their highest frequency \citep[e.g.,][]{Sobczak2000,Homan2001,Homan2005}.

\target is a new X-ray transient discovered by {\it MAXI}/GSC on 2019 January 26 \citep{Yatabe2019}. {\it Swift}/XRT measured the source location at ${\rm RA=13^{h} 48^{m} 12\fs7}$ and ${\rm DEC=-63\degr 16\arcmin 26\farcs 8}$, with an error radius of $1.7\arcsec$ (90\% confidence, \citealt{Kennea2019}). An optical counterpart was identified by \citet{Denisenko2019} and confirmed by \citet{Kennea2019}.
A radio counterpart was reported by \citet{Russell2019A}. 
MAXI J1348$-$630's multi-wavelength properties strongly suggest that the source is a black hole candidate (BHC) in a binary system \citep{Carotenuto2019,Denisenko2019,Jana2019,Kennea2019,Russell2019A,Sanna2019,Yatabe2019,Belloni2020}. 
In this work, we report results from the {\it Neutron Star Interior Composition Explorer} \citep[{\it NICER},][]{Gendreau2016} observations of MAXI J1348$-$630, which support the BHC interpretation.


\section{Observations and Data Analysis}
\label{sec:data}

The \nicer X-ray Timing Instrument (XTI) comprises an array of 56 co-aligned concentrator X-ray optics \citep{Gendreau2016}. Each optic is paired with a single-pixel silicon drift detector working in the 0.2--12 keV band. Presently, 52 detectors are working.  We excluded data from detectors \#14 and \#34, as they occasionally show episodes of increased electronic noise. The remaining 50 detectors provide a peak effective area of $\sim1800~\rm{cm}^2$ at 1.5 keV.

\begin{figure*}
\begin{center}
\resizebox{2\columnwidth}{!}{\rotatebox{0}{\includegraphics[clip]{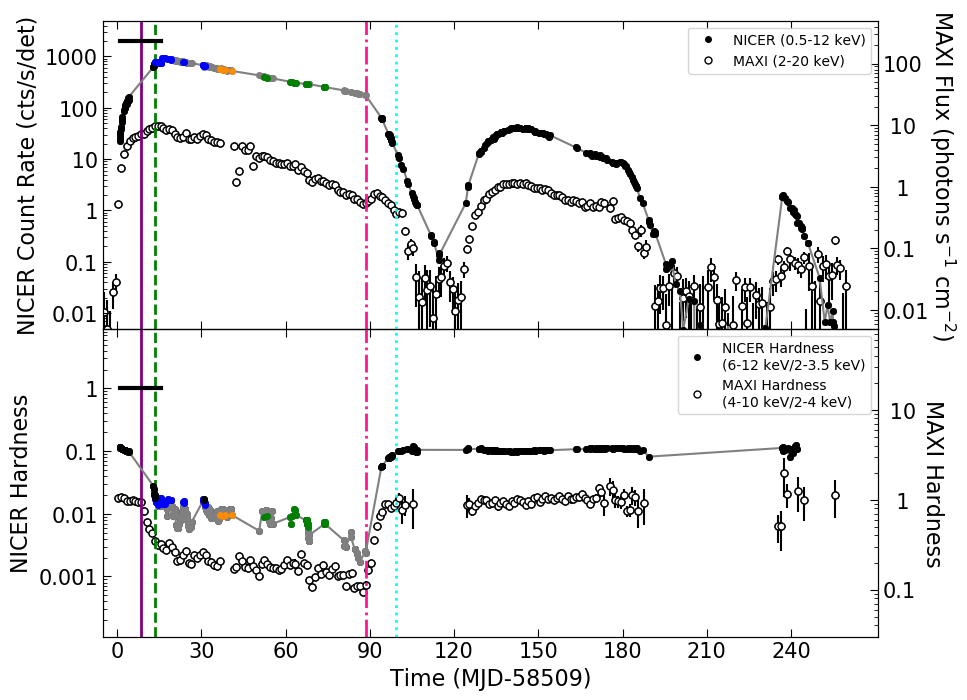}}}
\end{center}
\caption{Upper panel: Light curves of \target in the \nicer 0.5--12 keV (filled circles) and \maxi 2--20 keV (black open circles) energy bands. Lower panel: Evolution of the \nicer hardness (6--12 keV/2--3.5 keV) and \maxi hardness (4--10 keV/2--4 keV) during the outburst. In our analysis, we divided each \nicer observation into several continuous data segments, typically separated by several ksec gaps due to the orbit of the ISS. Here each point represents a single data segment. For clarity, in the lower panel we removed points where the hardness was unconstrained (i.e. large errors) due to the low count rate observed. The time origin is MJD 58509, the date when \maxi discovered the source. The purple solid line represents the time (MJD 58517) when the source starts the hard-to-soft transition. The green dashed line represents the time (MJD 58522) when the source enters the SIMS. The pink dash-dot line represents the time (MJD 58597) when the source starts the soft-to-hard transition. The cyan dotted line represents the time (MJD 58608) when the source goes back to the HS at the end of the outburst. Black points mark PDS with strong band-limited noise and sometimes a type-C QPO. Type-B and A QPOs are marked with blue and orange, respectively. Grey points mark PDS with little power. Green points mark PDS with relatively strong band-limited noise and sometimes a $\sim18$ Hz QPO observed in the soft state. See Fig. \ref{fig:pds} and corresponding text below for details. The black horizontal line marks \nicer data segments observed at a modest pointing offset before the source position was refined.}
\label{fig:lc}
\end{figure*}

Several hours after {\it MAXI}/GSC detected MAXI J1348$-$630, \nicer started  a near-daily monitoring campaign. We used all  \nicer data obtained between 2019 January 26 and 2019 October 8 (ObsIDs 1200530101--1200530128 and 2200530101--2200530231), including the following special cases: during ObsIDs 1200530101--1200530109 (MJD 58509$-$58524.8), before precise coordinates were available, \target was mistakenly observed with a pointing offset of 2.2 arcmin (${\rm RA}=13^{\rmn h} 47^{\rmn m} 55^{\rmn s}$ and ${\rm DEC}=-63\degr 15\arcmin 34\arcsec$); during ObsIDs 1200530106--2200530109 (MJD 58521--58560), some of the detectors were switched off to prevent telemetry saturation and accommodate the high source flux; we omitted detectors \#10--17 in ObsIDs 2200530169--2200530181 (MJD 58672--58687) owing to incorrect timestamps caused by a temporary instrument anomaly.

All data were processed using {\sc heasoft} version 6.26 and {\sc nicerdas} version 6.0. 
We used the latest calibration files obtained from the standard CALDB release for {\it NICER}, downloaded from NASA's High Energy Astrophysics Science Archive Research Center (HEASARC).
The gain calibration file was ``nixtiflightpi20170601v003\_optmv7he.fits", and the spectral response consisted of the detector redistribution matrix file ``nicer\_v1.02.rmf" and the optics auxiliary response file ``nicer-arf-consim135o-array$NN$.arf", where $NN$ is the number of active detectors\footnote{In some observations, some of the detectors were switched off. The corresponding auxiliary response files are the combination from all the active detectors.}.
We applied the standard filtering criteria to obtain the cleaned event files, i.e. time intervals requiring pointing stability $<54\arcsec$, bright Earth limb angle $>30\degr$, dark Earth limb angle $>15\degr$, and outside the South Atlantic Anomaly. Furthermore, we excluded times when overshoot count rate was larger than 1.0 counts per second per detector (cts/s/det) to remove strong background flare-ups. 
%
The background was calculated with the ``nibackgen3C50'' tool provided by the \nicer team. Individual \nicer observations have durations from $\sim1$ to $\sim80$ ksec, and are composed of one to multiple data segments of no more than 2--3 ksec gap-free data. Within the timescale of an observation, the background can show significant changes. Therefore, we divided each observation into several continuous data segments, typically separated by several ksec gaps due to the orbit of the ISS.
The total net exposure is $\sim274$ ksec.

We produced a light curve with 1-s binning in the 0.5--12 keV energy band for each data segment using XSELECT. In order to calculate the hardness ratio, we extracted light curves in the 2--3.5 keV band and the 6--12 keV band. We then applied background correction for each light curve, and divided the net count rate by the number of active detectors. The \nicer hardness is defined as the ratio between the 6--12 keV count rate and the 2--3.5 keV count rate. We then calculated the average count rate and hardness ratio for each data segment. For comparison, we also obtained 1-day bin \maxi light curves from the full 2--20 keV band as well as the 2--4 keV and 4--10 keV bands, through the \maxi on-demand Web interface\footnote{\url{http://maxi.riken.jp/pubdata/v6m/J1348-632/index.html}}. The \maxi hardness is defined as the ratio between the 4--10 keV rate and the 2--4 keV rate.

For the timing analysis, we produced an average PDS in the 0.5--12 keV energy band for each data segment. We used 16-s long intervals and 1/8192-s time resolution, so that the Nyquist frequency is 4096 Hz. The PDS were Leahy normalized \citep{Leahy1983} and converted to squared fractional rms \citep{Belloni1990}. The contribution due to the photon counting noise was subtracted. Following \citet{Belloni2002}, we fitted a selected sample of PDS with a model consisting of a sum of Lorentzian functions  (see section \ref{sec:QPO} for details).
We found spurious timing signatures (mainly below 0.5 Hz) in the PDS of the early observations made with a pointing-coordinate offset; these signatures are due to pointing jitter\footnote{See \url{https://heasarc.gsfc.nasa.gov/docs/nicer/data_analysis/nicer_analysis_tips.html} for details.} and the features are particularly significant  in ObsIDs 1200530107--1200530109, when the source was bright. 
To minimize the effect of these signatures, we chose the 0.5--64 Hz range for the total fractional rms estimation. We also calculated the absolute rms by multiplying fractional rms by net count rate \citep{Munoz2011}.

We extracted a background-subtracted energy spectrum for each data segment observed with a precise coordinate (i.e. we excluded all data between MJD 58509 and 58524.8). The spectra were corrected for the oversampling and rebinned to have at least 30 counts per bin. Instrumental residuals are seen at low energy band \citep[see][for detail]{Ludlam2018}. We therefore added a systematic error of 3\% and 0.5\% below and above 1.5 keV, respectively. The resulting spectra were analyzed using XSPEC version 12.10.1 in the 0.6--10 keV band.

\section{Results}
\label{sec:results} 

\subsection{Light curve and hardness}

In the upper panel of Fig. \ref{fig:lc} we show the \nicer (0.5--12 keV) and \maxi (2--20 keV) light curves of the outburst. Each data point represents the average for a single data segment.
At the beginning of the outburst, the source exhibited a fast ($\sim15$ days) rise with a peak \nicer count rate (0.5--12 keV) of $\sim950$ cts/s/det on day 16 (MJD 58525.2). It is important to mention that the source flux was underestimated due to the pointing offset\footnote{It is very complicated to estimate how much the flux was affected by the pointing offset and therefore we ignored the effect as it  does not affect the conclusions of this paper.} in observations before day 15 (MJD 58524.8). The peak count rate in the \maxi 2--20 keV band was $\sim9.6~\rm{photons~s^{-1}~cm^{-2}}$ ($\sim2.7$ Crab). 
Following the peak, the source flux decayed exponentially until day 88 (MJD 58597), when the source flux started to decrease at a higher rate. The source was undetected on day 106 (MJD 58615) with {\it MAXI}.
Starting on day 125 (MJD 58634), the source started a new outburst (or reflare). The second peak is much fainter than the first one, with a peak flux of $\sim0.3$ Crab ($\sim1.2~\rm{photons~s^{-1}~cm^{-2}}$ in the \maxi 2--20 keV band). The reflare profile is very similar to that of the main outburst, showing a fast rise and an initial slow decay followed by a faster decay. \target was undetected again around day 191 (MJD 58700).
Starting around day 232 (MJD 58741), the source started another reflare with a peak flux of $\sim0.03$ Crab ($\sim0.1~\rm{photons~s^{-1}~cm^{-2}}$ in the \maxi 2--20 keV band). \nicer observations did not sample the rise of this reflare, with a data gap of 6 days between non-detection and the peak.

In the lower panel of Fig. \ref{fig:lc}  we show the \nicer  (6--12 keV/2--3.5 keV)  and \maxi  (4--10 keV/2--4 keV) hardness ratios. The trends shown by \nicer and \maxi are very similar. During the first $\sim8$ days, the hardness remained more or less constant. After day 8 (MJD 58517), the \maxi hardness started to decrease rapidly, probably indicating a (hard-to-soft) state transition. The fast decay lasted $\sim5$ days. After day 13 (MJD 58522), the hardness decreased slowly until day 88 (MJD 58597), when the hardness started increasing to reach a level similar to that of the beginning of the outburst. During the reflares, the hardness stayed relatively constant, at a similar level to that of the hard states exhibited in the beginning and at the end of the main outburst. 

\begin{figure}
\begin{center}
\resizebox{1\columnwidth}{!}{\rotatebox{0}{\includegraphics{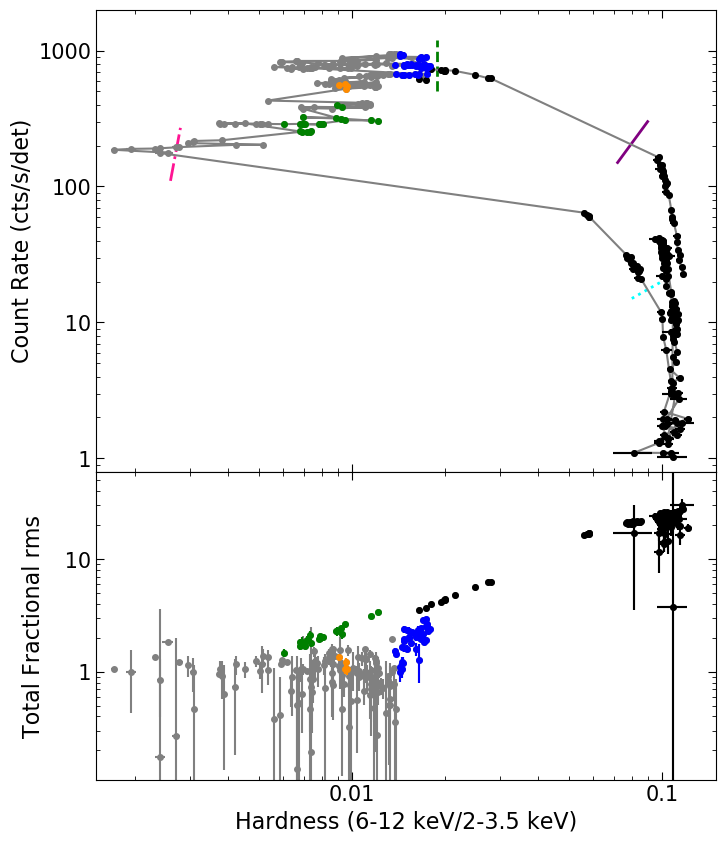}}}
\end{center}
\caption{\nicer hardness-intensity diagram (HID, upper panel) and hardness-rms diagram (HRD, lower panel) of the outburst. The total fractional rms is calculated in the 0.5--64 Hz frequency range. Symbols are the same as those in Fig. \ref{fig:lc}.}
\label{fig:hid}
\end{figure}

\begin{figure}
\begin{center}
\resizebox{1\columnwidth}{!}{\rotatebox{0}{\includegraphics{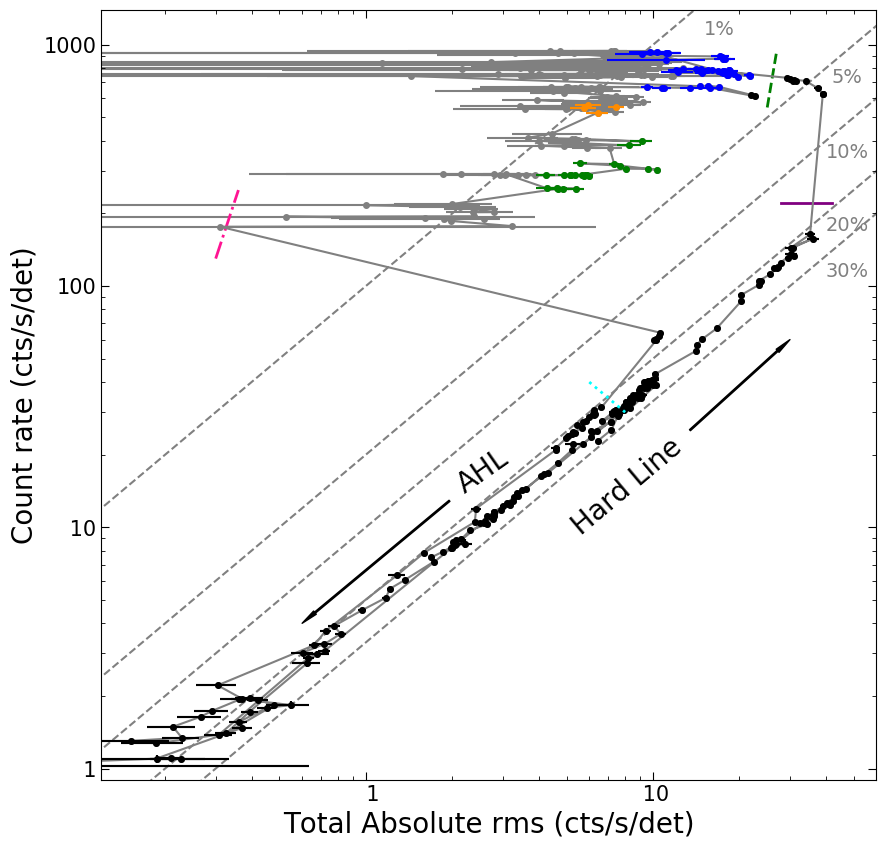}}}
\end{center}
\caption{\nicer absolute rms-intensity diagram (RID) of the outburst. Symbols are the same as those in Fig. \ref{fig:lc}. The gray dotted lines represent the 1, 5, 10, 20 and 30 percent fractional rms levels.}
\label{fig:rid}
\end{figure}

\subsection{Averaged spectral and X-ray variability evolution}

In the top panel of Fig. \ref{fig:hid} we show the evolution of the source in the hardness-intensity diagram (HID); in the bottom panel we show how the 0.5--64~Hz averaged fractional rms amplitude evolves with hardness (HRD). 
Fig. \ref{fig:hid} shows that \target traced a  `q' pattern in the HID, evolving in an anticlockwise direction. This is very similar to what is generally seen from BHCs \citep[e.g.][]{Belloni2005,Homan2005,Kalamkar2011,Motta2011,Zhang2015}. Based on the location in the HID, 
the outburst started from the right vertical branch, corresponding to the canonical HS. 
Between days 5--11 (MJD 58514--58520), there were no \nicer observations. On day 12 (MJD 58521), the source was already in the upper horizontal branch, where the HIMS and SIMS are usually expected. After that, the source gradually evolved to the left in the HID (i.e. to softer spectra), as the source intensity also started to decrease.
At the end of the main outburst, around day 88 (MJD 58597), the source started to undergo a transition to harder colors, suggestive of a typical BH soft-to-hard transition as the source returned to the HS around day 99 (MJD 58608). Unfortunately, most of the soft-to-hard transition was not sampled with \nicer observations.

\begin{figure*}
\begin{center}
\resizebox{2.1\columnwidth}{!}{\rotatebox{0}{\includegraphics{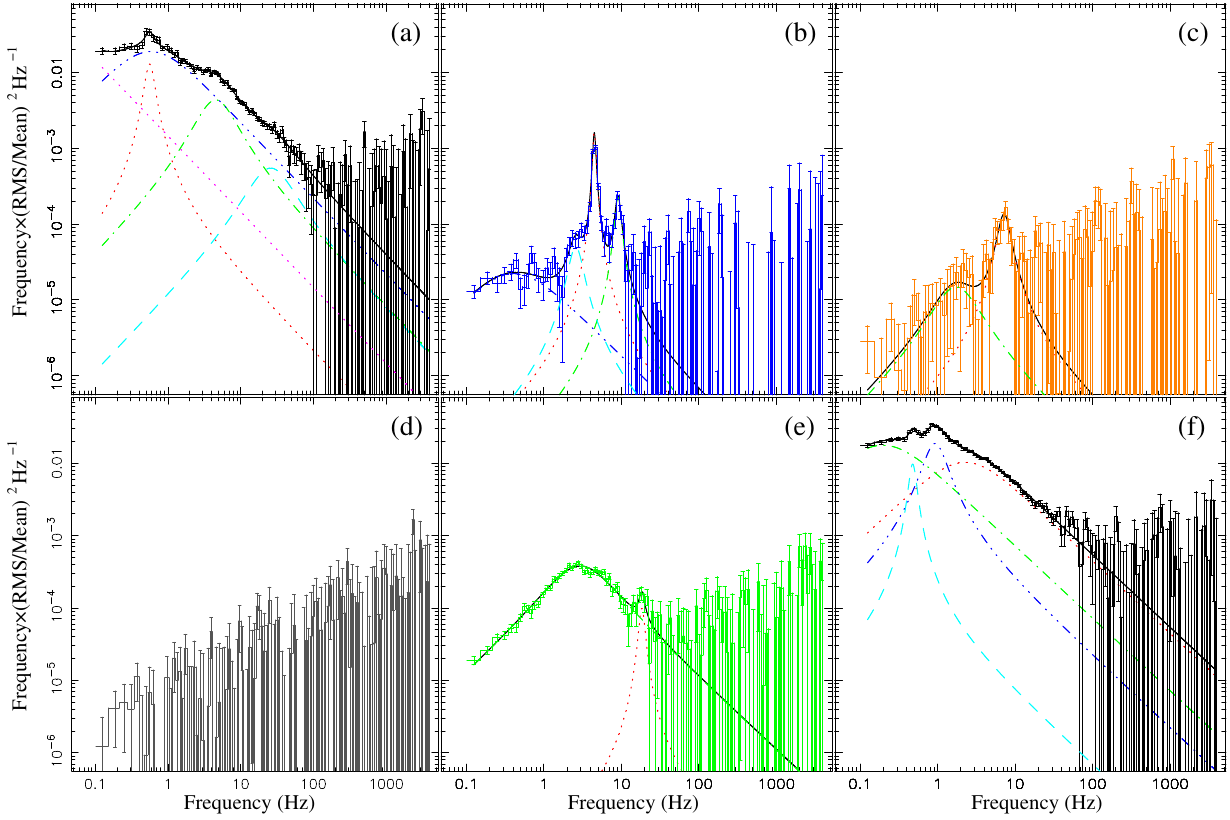}}}
\end{center}
\caption{Representative power spectra at different phases of the outburst.   The power spectra are calculated in the 0.5--12 keV band and fitted with a model composed of multiple Lorentzian functions. The main properties of the power spectra are listed in Tab. \ref{tab:fit}. The axis scales are consistent to compare the rms level of different types of PDS.}
\label{fig:pds}
\end{figure*}

The bottom panel of Fig. \ref{fig:hid} shows that the 0.5--64~Hz averaged fractional rms decreases as the hardness ratio decreases. This relation is tighter at hardness $>0.02$. At hardness lower than $\sim0.02$, we observe a drop of the rms amplitude. This drop corresponds to day 13 (MJD 58522.6), and for BHC is usually related to the transition between HIMS and SIMS \citep[see, e.g.,][and references therein]{Belloni2011}.

During the reflares, the source evolved along the right vertical branch in the HID (Fig. \ref{fig:hid}), suggesting that it stayed in the hard state only. The fractional rms amplitude was between 20\% and 30\%, similar to that at the beginning of the main outburst. This behavior is reminiscent of the so-called ``failed outbursts", where the outburst of the BHC does not show any transition to the soft state \citep[e.g.,][]{Sturner2005,Capitanio2009,Del2016}.


In Fig. \ref{fig:rid} we show how \target evolves in the absolute rms-intensity diagram (RID). The system here also traced an anticlockwise hysteresis pattern, similar to what has been seen in other BHCs. At the beginning of the outburst, we found a tight correlation between the absolute rms and flux. Such a linear relation has been observed in many other BHTs, and is commonly known as the ``hard line'' \citep[HL,][]{Munoz2011}.
After the gap between days 5 and 11 (MJD 58514 -- 58520), the absolute rms deviated from the HL on day 12 (MJD 58521), already crossing the 10\% fractional rms line, and then moving horizontally to the left of the diagram.
Then, \target evolved to lower intensity, following approximately the $1\%$ fractional rms line.
%
Around day 99 (MJD 58608), simultaneously with the transition observed in the HID, we observe that the fractional rms amplitude increased to levels seen at the beginning of the outburst. This is consistent with the source returning to the HS, and evolving along the so-called ``adjacent hard line" \citep[AHL,][]{Munoz2011}.
During the reflares, the source moved upwards along the HL as the flux rose, and back along the same track during the decay.

\begin{table*}
\centering
\caption{Properties of the power spectra shown in Fig. \ref{fig:pds}. The second column lists the time used to produce the PDS. Columns 3--5 list the centroid frequency, $Q$ factor and fractional rms of the QPO and its harmonics shown in the PDS. The last column lists the 0.5--64 Hz fractional rms.}
\label{tab:fit}
\begin{tabular}{lccccc}
\hline\hline
Panel  &      Date       &     $\nu_{0}$   &      $Q$        &      QPO rms     &  0.5--64 Hz rms \\
       &     (MJD)       &       (Hz)      & ($\nu_{0}/\rm FWHM$)   &        (\%)      &      (\%) \\
\hline
(a)  &  58512.1--58512.9 & $0.56 \pm 0.02$ & $2.88 \pm 1.24$ & $8.13 \pm 1.59$  & $22.47 \pm 0.18$\\
\hline
(b)  &  58527.0--58528.2 & $4.49 \pm 0.01$ & $9.88 \pm 1.04$ & $1.58 \pm 0.03$  & $1.94 \pm 0.05$\\
     &                   & $9.07 \pm 0.11$ & $5.04 \pm 0.96$ & $0.82 \pm 0.05$  & -\\
     &                   & $2.59 \pm 0.16$ & $2.22 \pm 0.74$ & $0.57 \pm 0.08$  & -\\
\hline
(c)  &  58547.4          & $7.26 \pm 0.24$ & $2.66 \pm 0.93$ & $0.85 \pm 0.10$  & $1.05 \pm 0.11$\\
\hline
(d)  &  58594.5--58594.8 &       -         &       -         &       -          & $0.85$\\
\hline
(e)  &  58571.4--58572.7 & $18.47 \pm 0.80$ & $5.23 \pm 2.67$ & $0.54 \pm 0.09$ & $2.93 \pm 0.03$\\
\hline
(f)  &  58655.1--58656.0 & $0.93 \pm 0.02$  & $1.35 \pm 0.21$ & $13.69 \pm 1.18$ & $24.54 \pm 0.10$\\
     &                   & $0.48 \pm 0.01$  & $4.06 \pm 2.29$ & $5.97 \pm 0.98$  &  -\\
\hline
\end{tabular}
\end{table*}

\subsection{Broad band noise and quasi-periodic oscillations}
\label{sec:QPO}

The X-ray variability of \target evolved during the outburst. In Fig. \ref{fig:pds} we show six representative PDS at different phases of the outburst. We have plotted each type of PDS with the colors used in Figures \ref{fig:lc} through \ref{fig:rid}, where the panels a, b, c, d and e correspond to black, blue, orange, gray, and green, respectively; panel f shows a PDS taken during the reflare, which is similar to that of panel a.

During the rise of the main outburst, the PDS are dominated by a strong ($\sim20$--30\% rms) band-limited noise component and sometimes a low-frequency QPO (Fig. \ref{fig:pds}a). The characteristic frequency of the QPO increases from $\sim0.2$ Hz on MJD 58510 to $\sim5$ Hz on MJD 58522. 
At hardness $\sim0.02$ (MJD 58522.6), we found a drop of the fractional rms amplitude. 
At the same time, the strong band-limited noise is replaced by a weak broadband noise component. A sharp, narrow ($Q>6$) QPO is detected in the PDS, together with its second harmonic. The QPO appears in a narrow hardness range with a stable frequency around 4 Hz. Fig. \ref{fig:pds}b shows a representative example. The QPO sometimes disappears and reappears within a timescale of one day. This PDS is reminiscent of the type-B QPOs observed in BHC.
%
In most cases after MJD 58540.3, the PDS reveals little variability ($\sim1\%$ rms, Fig. \ref{fig:pds}d). However, we found relatively strong band-limited noise between MJD 58570--58576, sometimes with a weak but significant ($\sim3 \sigma$) QPO around 18 Hz (Fig. \ref{fig:pds}e). In addition, in four cases around hardness $\sim0.01$, we detected a broader QPO (Fig. \ref{fig:pds}c) than that in Fig. \ref{fig:pds}b around $\sim7$ Hz, but still accompanied by weak broadband noise. Around MJD 58562, we observed PDS similar to that in Fig. \ref{fig:pds}b with a less significant ($<3\sigma$) peaked Lorentzian around 2 Hz. 
This QPO becomes significant if data from several segments are averaged together (e.g. we detect a single trial $3.5\sigma$ $2.3\pm0.3\%$ fractional rms amplitude QPO with $Q=2.4\pm0.8$ at $1.78\pm0.06$ Hz when averaging data from ObsIDs 2200530110-1 and 2200530113 in the 3-12 keV range).
%
When the source transitioned back to the right side of the HID (consistent with the HS), the PDS is characterized again by the band-limited noise component; sometimes we found weak QPOs with characteristic frequency decreasing from $\sim3$ Hz (MJD 58603) to $\sim0.1$ Hz (MJD 58612).

The PDS during the reflares are similar to that observed at the end of the main outburst, i.e, the PDS are characterized by a strong band-limited noise component and sometimes a low-frequency QPO with frequencies between $\sim0.1$ Hz and $\sim 1$ Hz. Fig \ref{fig:pds}f shows an example PDS taken at the peak of the first reflare (MJD 58655).

Based on comparisons with previous work on other sources \citep[e.g.,][]{Casella2004,Rodriguez2004,Belloni2005,Homan2005b,Motta2011}, the power spectral evolution we observed is consistent with that of BHCs. Panels a and f in Fig. \ref{fig:pds} exhibit the so-called type-C QPO PDS, while that in panel b resembles a type-B. Under the BHC interpretation, panels c and d in Fig. \ref{fig:pds} show type-A QPOs and a soft state PDS, respectively. BHCs sometimes also exhibit the type of PDS we show in Fig. \ref{fig:pds}e \citep[e.g.,][]{Homan2001,Homan2005b,Motta2012}.

\begin{figure}
\begin{center}
\resizebox{1.02\columnwidth}{!}{\rotatebox{0}{\includegraphics{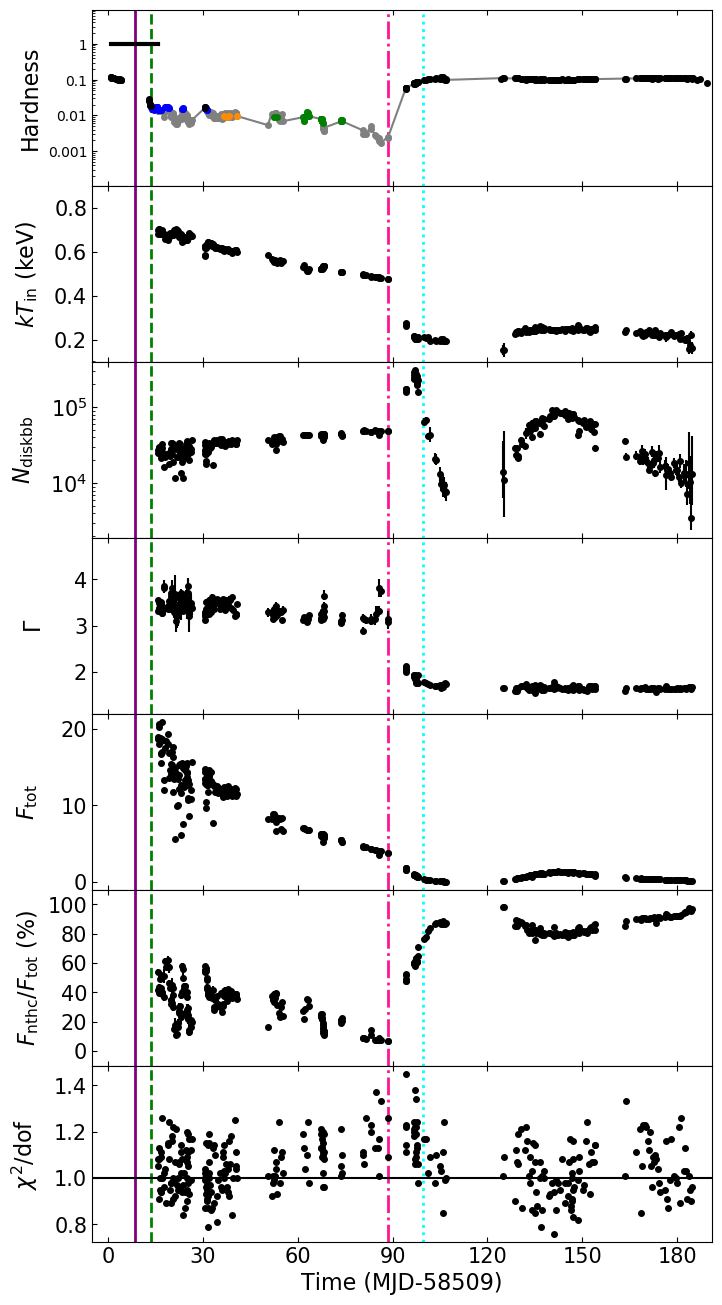}}}
\end{center}
\caption{Evolution of the main spectral parameters. The continuum was fitted with the model \textsc{tbnew*(diskbb+nthcomp)} in the 0.6--10 keV band. In the top panel we show the hardness ratio for comparison. From the second panel to the bottom: disk temperature at the inner radius ($kT_{\rm in}$) in keV, \textsc{diskbb} normalisation ($N_{\rm diskbb}$), photon index ($\Gamma$), unabsorbed total flux ($F_{\rm tot}$) in units of $10^{-8}$ erg cm$^{-2}$ s$^{-1}$, the contribution of the Comptonised component flux to the total flux ($F_{\rm nthc}/F_{\rm tot}$), and the reduced $\chi^{2}$. We do not plot the data after day 187 (MJD 58596), since the disk component was not required by the spectral fits and the spectra can be fitted well with a single power-law with $\Gamma\sim1.6$.}
\label{fig:spectral_para}
\end{figure}

\subsection{Energy Spectra}
\label{sec:ep}

\begin{figure}
\begin{center}
\resizebox{1\columnwidth}{!}{\rotatebox{0}{\includegraphics{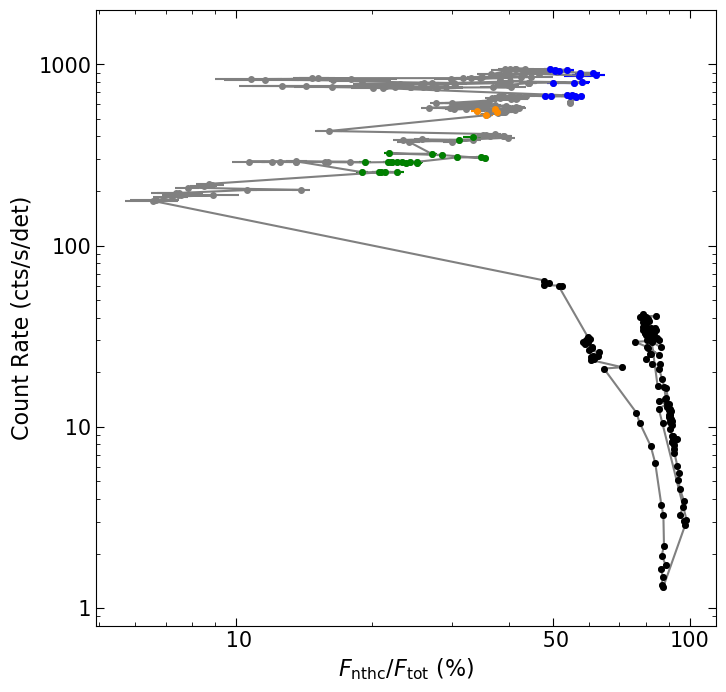}}}
\end{center}
\caption{\nicer 0.5--12 keV count rate as a function of $F_{\rm nthc}/F_{\rm tot}$. Symbols are the same as those in Fig. \ref{fig:lc}.}
\label{fig:FID}
\end{figure}

\begin{figure}
\begin{center}
\resizebox{1\columnwidth}{!}{\rotatebox{0}{\includegraphics{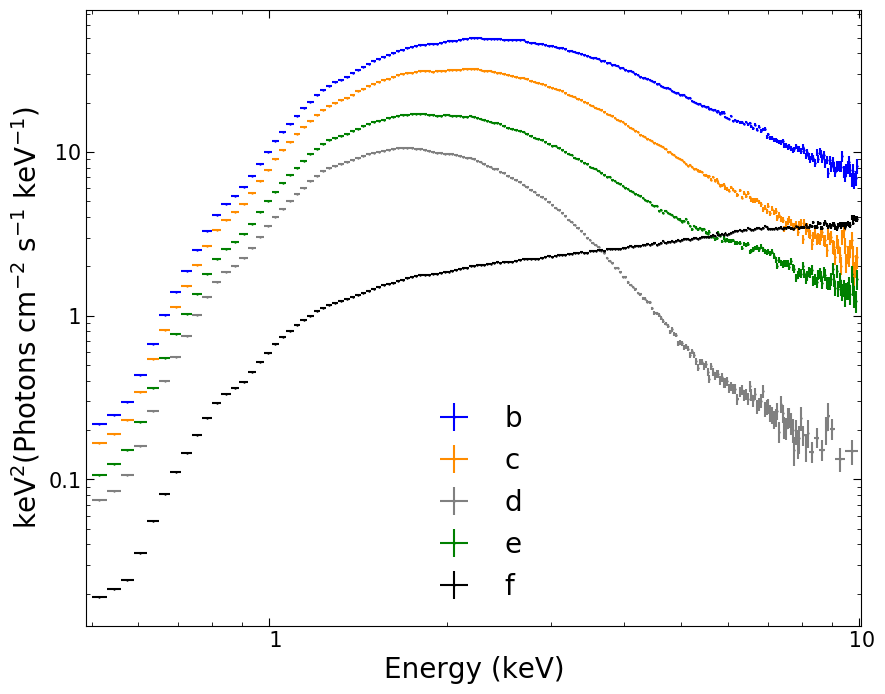}}}
\end{center}
\caption{Representative energy spectra corresponding to different types of PDS shown in panels b-f in Fig. \ref{fig:pds}. The energy spectrum during the time used to produce the PDS in panel a is not shown due to the imprecise pointing of the instrument. The energy spectra corresponding to panels b-f in Fig. \ref{fig:pds} are extracted from the same time used to produce the PDS.}
\label{fig:ep}
\end{figure}

The energy spectra were fitted with a two-component model consisting of a multi-temperature blackbody component (\textsc{diskbb}: \citealt{Mitsuda1984}) and a thermally Comptonised continuum (\textsc{nthcomp}: \citealt{Zdziarski1996,Zycki1999}). The seed photon temperature of the Comptonised component, $kT_{\rm bb}$, was linked to the inner disk temperature of the disk component, $kT_{\rm in}$. Since the high energy cutoff of the Comptonised component is outside the \nicer energy range, we fixed the electron temperature at 1000 keV. The absorption along the line of sight was modelled using \textsc{tbnew}, which is an improved version of the X-ray absorption model \textsc{tbabs}\footnote{\url{https://pulsar.sternwarte.uni-erlangen.de/wilms/research/tbabs/}}. The oxygen and iron absorption abundances were fixed at the solar value. The fits are more sensitive to the hydrogen column density, $N_{\rm H}$, when the spectra are soft. We then fixed $N_{\rm H}$ at $0.64\times10^{22}~{\rm cm}^{-2}$, the value obtained from the best-fits of the spectra during the soft state.
A weak Fe-K emission feature was marginally observed in most of the spectra. A \textsc{gaussian} line was added if the addition of this component improves the fits significantly by $\Delta\chi^{2} \gtrsim 30$ (corresponding to an F-test\footnote{It is important to mention that F-test could underestimate or overestimate the significance of a narrow line feature in some cases \citep{Protassov2002}; however, we do not investigate this further as variations in the true significance of the Fe-K line  does not affect the evolution of the main spectral parameters shown in Fig. \ref{fig:spectral_para}.} probability of $\lesssim10^{-6}$ for an average of $\sim220$ dof) for three fewer dof. The centroid of the Fe-K line was allowed to vary between 6.3 and 6.7 keV; the width of the Fe-K line was constrained between 0.1 and 1 keV. In all cases, the contribution of the Fe-K line flux to the total flux is less than 1.3\%. 
Extra residuals were observed below 3 keV after day 94 (MJD 58603) which is possibly due to instrumental effects. We added another \textsc{gaussian} component to account for these residuals if the addition of this component improved the fits by $\Delta\chi^{2} \gtrsim 30$. The centroid of the line was fixed at 1 keV; the width of the line obtained remains more or less constant at 0.9 keV. The effect of the addition component on $kT_{\rm in}$ and photon index, $\Gamma$, is typically less than 0.03 keV and 0.1, respectively. 
After day 187 (MJD 58596), the disk component was not required by the spectral fits, and the spectra can be fitted well with a single power-law with $\Gamma\sim1.6$.
The average reduced $\chi^{2}$ from the best fits was 1.1. 
We also calculated the unabsorbed flux of different components using \textsc{cflux} in the 0.6--10 keV band. In this work, we will focus on the evolution of the disk and Comptonised components during the outburst. A more detailed spectral analysis will be presented in a separate paper.

In order to test the robustness of our spectral parameters, we also tried another continuum model \textsc{tbnew*(simpl*diskbb)}, in which a fraction of the seed photons from the accretion disk is up-scattered into a power-law component  (\textsc{simpl}: \citealt{Steiner2009}). The evolution of $kT_{\rm in}$ and $\Gamma$ are consistent between the two models we used.
Our results are also consistent with those recently presented by \citet{Tominaga2020}, who fitted the 2--20 keV MAXI/GSC spectra of  MAXI J1348$-$630 with the model \textsc{tbabs*(simpl*diskbb)}. We note, however, that these authors could not constrain the photon index during the soft state (and therefore fixed it at 2.5) and could not detect the disk component during the reflares in the MAXI energy band.

%

In Fig. \ref{fig:spectral_para} we show the evolution of the hardness ratio and the main spectral parameters from the model (hereafter model 1) \textsc{tbnew*(diskbb+nthcomp)}. From the top to the bottom panels, we plot the hardness, inner disk temperature ($kT_{\rm in}$), \textsc{diskbb} normalisation ($N_{\rm diskbb}$), photon index ($\Gamma$), total unabsorbed flux ($F_{\rm tot}$), the contribution of the Comptonised component flux to the total flux ($F_{\rm nthc}/F_{\rm tot}$), and the reduced $\chi^2$. In Fig. \ref{fig:FID} we plot the intensity versus $F_{\rm nthc}/F_{\rm tot}$. In Fig. \ref{fig:ep}, we show the energy spectra corresponding to the PDS in Fig. \ref{fig:pds} b-f. The energy spectrum during the time used to produce the PDS in panel a is not shown due to the imprecise pointing of the instrument. 
For completeness, we show the evolution of the spectral parameters from the model (hereafter model 2) \textsc{tbnew*(simpl*diskbb)} in Fig. \ref{fig:simpl} (Appendix). We found that:
\begin{itemize}
\item Between day 16--88 (MJD 58525--58597), the spectra gradually softened with a decrease in total flux. During this period, the photon index underwent a slight decrease from $\Gamma\sim3.5$ to $\sim3$ along with a decrease in inner disk temperature from $kT_{\rm in}\sim0.7$ to $\sim0.5$ keV. The disk component became more dominant, and the contribution of the Comptonised component flux to the total flux ($F_{\rm nthc}/F_{\rm tot}$) decreased from $\sim60\%$ to $<10\%$. 
\item After that the source started a soft-to-hard transition until day 99 (MJD 58608). During the soft-to-hard transition, the inner disk temperature decreased to $\sim0.2$ keV and the photon index decreased to $\sim1.7$, while the contribution of the Comptonised component flux to the total flux ($F_{\rm nthc}/F_{\rm tot}$) increased to $>70\%$.
\item After the source returned back to the hard state around day 99 (MJD 58608) and during the reflares, the spectra are dominated by the Comptonised component with $F_{\rm nthc}/F_{\rm tot}\gtrsim80\%$. The photon index remains approximately constant at $\sim1.6$, and the inner disk temperature is around 0.2 keV.
\item The spectrum corresponding to the PDS shown in Fig. \ref{fig:pds}a is not studied due to the imprecise pointing of the instrument. Based on its position in Fig. \ref{fig:hid}--\ref{fig:rid} and the timing characteristics, it should have a similar spectral shape as the spectrum corresponding to the PDS shown in Fig. \ref{fig:pds}f. The PDS in Fig. \ref{fig:pds}f corresponds to typical hard-state spectra, which are dominated by the Comptonised component ($F_{\rm nthc}/F_{\rm tot}\gtrsim80\%$) with a photon index of $\Gamma\sim1.6-1.7$;
when PDS in Fig. \ref{fig:pds}b were observed, the spectra have a typical inner disk temperature and photon index of $kT_{\rm in}\sim0.65$ keV and $\Gamma\sim3.4$. The contribution of the Comptonised component flux to the total flux ($F_{\rm nthc}/F_{\rm tot}$) is $\sim50-60\%$;
the spectra corresponding to the PDS shown in Fig. \ref{fig:pds}c are similar to that shown in Fig. \ref{fig:pds}b, but with a slightly lower inner disk temperature ($kT_{\rm in}\sim0.60$ keV) and a lower $F_{\rm nthc}/F_{\rm tot}\sim35\%$;
the PDS in Fig. \ref{fig:pds}d corresponds to soft-state spectra which are dominated by a strong disk component with $F_{\rm nthc}/F_{\rm tot}$ typically $<20\%$;
when PDS in Fig. \ref{fig:pds}e were detected, the spectra are characterised by an inner disk temperature of $kT_{\rm in}\sim0.52$ keV and a photon index of $\sim3.2$ with $F_{\rm nthc}/F_{\rm tot}\sim20-35\%$.
\end{itemize}

The inner disk radius, $R_{\rm in}$, can be inferred from the normalisation of the \textsc{diskbb} component ($N_{\rm diskbb}$). 
From Fig. \ref{fig:spectral_para} and Fig. \ref{fig:simpl}, it can be seen that the evolution of $N_{\rm diskbb}$ during the soft state of the main outburst depends on the model used. When using ``model 1" (Fig. \ref{fig:spectral_para}), $N_{\rm diskbb}$ slightly increases as the outburst evolves, suggesting that the disk was receding as the total flux decreases. 
However, when using ``model 2" (Fig. \ref{fig:simpl}), $N_{\rm diskbb}$ remains more or less constant during the soft state. This is consistent with our expectation for the standard evolution of black hole binaries in the soft state \citep[e.g.,][]{Done2007,Steiner2010,Munoz2011}, and also consistent with the results shown in \citet{Tominaga2020}, where the constant $R_{\rm in}$ is thought to correspond to the innermost stable circular orbit (ISCO) of the black hole \citep[e.g.,][]{Steiner2010}.
The apparent difference $N_{\rm diskbb}$ between models 1 and 2 is expected given their implementations.  Both invoke Compton scattering and assume that the corona itself produces no intrinsic emission (i.e., bremsstrahlung and synchrotron are negligible), and so every photon in the Compton component should have originated in the disk.
When using \textsc{simpl}, the disk normalization is such that so that every X-ray photon from disk and power-law components is forced to originate in the same underlying (``seed'') disk.   $N_{\rm diskbb}$ accordingly tracks the seed disk radius \citep{Steiner2009}. However, in model 1, applying \textsc{nthcomp} \citep{Zdziarski1996} independent of the disk component is equivalent in practice to assuming a separate emission component (one distinct from the disk) produces the coronal flux, and here the disk flux normalization only describes the part which has not been scattered. Therefore, $N_{\rm diskbb}$ would {\em appear} to vary inversely with the optical depth of the corona.


Presently, the distance and inclination of \target are unknown. \citet{Tominaga2020} noted that \target is in the direction of the Galactic Scutum-Centaurus arm and (due to the low column density) may lie in front of it, at $\sim3-4$ kpc. Just as an example, if we assume the nearer distance of 3 kpc (also suggested by the large observed flux at the soft-to-hard transition) and an intermediate inclination of 60 degrees, the average $R_{\rm in}$\footnote{The value is corrected for the spectral hardening with a factor of $\kappa=1.7$ and the inner boundary condition with a factor of $\xi=0.41$ \citep{Kubota1998}.} in the  soft state using ``model 2" leads to $R_{\rm in} = 110\pm5$ km, which is consistent with the radius of ISCO for a non-spinning black hole of 12 M$_{\sun}$.

%

During the soft-to-hard transition and reflares, the evolution of $N_{\rm diskbb}$ is almost independent of any of the two models we used. During the transition, we found that $N_{\rm diskbb}$ undergoes a fast rise followed by a steep decay. Although this behaviour is contrary to simple expectations that the disk either simply recedes during the transition or remains at the ISCO with a change in coronal power, the sharply increasing then decreasing inferred radius was also observed in the soft-hard transition of black hole candidate XTE~J1817-330 by \citep{Gierlinski2008} and interpreted by those authors as being due to increasing coronal heating of the inner edge of a receding disk. 
As explained in section \ref{sec:ep} above,  after day 94 (MJD 58603) we had to include an additional \textsc{Gaussian} component to account for the extra residuals below 3 keV. The addition of the  \textsc{gaussian} component has a significant effect on $N_{\rm diskbb}$, but has little effect on the evolution of $kT_{\rm in}$ and $\Gamma$. Furthermore, we found during this period the  \textsc{gaussian}, $N_{\rm diskbb}$, and $N_{\rm H}$ become highly degenerate, but not enough to explain the unusual evolution of the disk normalisation.

During the first reflare, we found that $N_{\rm diskbb}$ first increased and then decreased, following the changes in flux. These results are inconsistent with the picture that the disk is moving inwards as the flux increases. One possibility is that our results reflect physical changes (e.g., due to changing coronal irradiation of the disk) that we were not able to detect before given the lack of low-energy coverage now provided by NICER. Another possibility is that the changes we see in $N_{\rm diskbb}$ are due to a combination of choice of model (the models we used here are relatively simple) and the lack of data above 10 keV (which do not allow us to constrain well all the model parameters, e.g., to properly account for reflection). This, together with evolution of $N_{\rm diskbb}$ during the soft-to-hard transition,  should be further explored using data covering a broader range of energies, i.e., using potential NICER+Astrosat or NICER+HXMT simultaneous observations.


\section{Discussion and Summary}

In this work, we present the first detailed analysis of the spectral evolution and timing properties of the newly discovered X-ray transient \target using \nicer observations. \target traces tracks in the HID, HRD, and RID which are typical of what is seen in BHTs \citep[see, e.g.,][and references therein]{Homan2005}. This interpretation is supported by the PDS evolution, and the fact that we detected Type-A, Type-B and Type-C QPOs \citep[see][]{Casella2004,Motta2011,Motta2016}. \target also underwent reflares at the end of the outburst. These reflares evolve similarly to the so-called ``failed outbursts", i.e. outbursts which are generally less luminous at their peak, and do not show any evidence of state transitions. 
Our results show that \target underwent an initial full outburst which showed all canonical states. Below in section \ref{sec:states} we summarize the different states, their characteristics, and their relation to recent results on investigations about \target at other wavelengths. In section \ref{sec:comp} we compare our results with that observed in other BHTs.

\subsection{Spectral states}
\label{sec:states}

\begin{itemize}
\item \textbf{Hard state} -- From MJD 58509 to MJD 58517, the source was in the HS, corresponding to the right vertical branch of the HID (Fig. \ref{fig:hid}). On MJD 58509, the {\it Swift}/XRT spectrum can be fitted by a power-law with $\Gamma=1.32\pm0.02$ \citep{Kennea2019}. 
During the HS, \target evolved along the hard line in the RID (Fig. \ref{fig:rid}). The corresponding PDS are dominated by band-limited noise and a QPO with variable frequency ranging from $\sim0.2$ Hz (MJD 58510) to $\sim0.7$ Hz (MJD 58513). We identify the power spectral shape and the type of QPOs as the so-called Type-C QPOs. The radio and optical flux also rise during this period \citep{Russell2019B,Russell2019A}. Our identification of the HS is supported by radio observations taken on MJD 58509.9 and MJD 58511.0. These observations show a flat or slightly inverted radio spectrum, suggesting that the radio emission comes from a compact jet \citep{Russell2019A}.

\item \textbf{Hard-intermediate state} -- Starting from MJD 58517 (purple solid line in Figs. \ref{fig:lc}--\ref{fig:rid}), the \maxi hardness drops rapidly with a clear spectral softening \citep{Nakahira2019}. During MJD 58517--58519, the power-law index obtained from {\it Swift}/XRT spectra increases from 1.68 to 2.25 \citep{Bassi2019}, which is evidence for a hard-to-soft transition. Unfortunately, most of the transition was not sampled with \nicer observations. The first \nicer observation in the HIMS was taken on MJD 58521.8, where the source had already left the hard line in the RID. During the transition, the fractional rms drops from $\sim20\%$ to $\sim5\%$, while the characteristic frequency of the type-C QPO increases to $\sim5$ Hz. The source only stays in the HIMS for $\sim5$ days. 
\item \textbf{Soft-intermediate state} -- On MJD 58522.6 (green dashed line in Figs. \ref{fig:lc}--\ref{fig:rid}), a sudden decrease in fractional rms amplitude was observed. At the same time, a narrow QPO with a stable frequency around 4 Hz and its second harmonic appear in the PDS, accompanied by a weak broadband noise component. These QPOs appear only in a narrow hardness range and a fractional rms of 1--5\%. Fast appearance and disappearance of the QPOs within a timescale of 1 day is observed. These transitions will be reported in more detail in a separate work. Based on these results, we identify these QPOs as type-B QPOs. The presence of type-B QPOs provides evidence that the source enters the SIMS \citep[a detailed analysis of the phase lags of the type-B QPOs has been recently reported by][]{Belloni2020}. The energy spectra during the SIMS have a significant contribution from both the disk and the Comptonised component ($F_{\rm nthc}/F_{\rm tot}\sim50-60\%$). The HIMS to SIMS transition is believed to be associated with the ejection of transient jets \citep{Fender2004}. Indeed, a bright radio flare was observed on MJD 58523 with MeerKAT \citep{Carotenuto2019}.
\item \textbf{Soft state} -- After the SIMS is reached, the fractional rms sometimes decreases to below 1\%. Based on the rms value, the source seems to enter the SS. We find QPOs with a characteristic frequency of $\sim7$ Hz in four data segments during this period. These QPOs are very broad ($Q<3$) and weak ($<1\%$ rms). Neither second harmonic nor subharmonic are observed. We identify these as type-A QPOs \citep[e.g.,][]{Casella2004,Belloni2011,Motta2016}. Also in some data segments during this period, we found the type of PDS shown in Fig. \ref{fig:pds}e with a relatively strong band-limited noise and sometimes a QPO at $\sim18$ Hz.  During the soft state, the spectra gradually softened as the source decay.
\item \textbf{Soft-to-hard transition and the end of the outburst} -- Starting from MJD 58597 (pink dash-dot line in Figs. \ref{fig:lc}--\ref{fig:rid}), the source experienced a final soft-to-hard transition with an increase in \maxi hardness. On MJD 58608 (cyan dotted line in Figs. \ref{fig:lc}--\ref{fig:rid}), the source went back to the HS and evolved along the AHL in the RID. 
The PDS are similar to that observed in the early HS. However, the spectral hardness is slightly softer in the final HS. 

After the main outburst, \target re-brightened again and experienced at least two reflares. The light curve profile of the first reflare is similar to that of the main outburst. However, the spectral evolution is quite different. During the reflares, the source stayed in the right vertical branch in the HID. The PDS are typical of the HS, which are dominated by strong band-limited noise. The energy spectra are dominated by the Comptonized component with a photon index of $\Gamma\sim1.6$. These results show that the source remained in the HS, never making a transition to the SS.
\end{itemize}

\subsection{Comparison with previous results}
\label{sec:comp}

The spectral evolution of \target during its main outburst is similar to what was previously observed in other BHTs \citep[see, e.g.,][]{Belloni2005,Homan2005,Kalamkar2011,Zhang2015}. The source went from the HS, through the HIMS and SIMS, into the SS in the outburst rise, and back to the HS in the outburst decay. 
The hard-to-soft transition was very fast. The source only stayed in the HIMS for $\sim5$ days, and then moved into the SIMS characterised by the appearance of the type-B QPOs. This is similar to the 2006 and 2010 outbursts of GX 339$-$4, where the source stayed in the HIMS for $\sim7$ days \citep{Motta2009,Motta2011}. However, in some outbursts of some sources, the HIMS lasts much longer ($>10$ days, e.g. \citealt{Munoz2011}). We note that in some cases, the systems can go back to the HIMS from the SIMS \citep{Belloni2011}. 
During the SS, the disk temperature is in the range of 0.5--0.7 keV, which is slightly lower than that observed in other BHTs \citep{Dunn2011}. For a standard accretion disk, the disk temperature is inversely related to the black hole mass ($T \propto M^{-1/4}$, \citealt{Shakura1973}). The lower disk temperature of \target suggests that it may harbour a relatively higher mass black hole. 
The photon index obtained from the fits during the SS ($\Gamma\sim3-3.5$) is higher than the $\Gamma\sim2.2-3$ that has been seen in previous works \citep[e.g.,][]{Remillard2006,Motta2009}. The higher values of $\Gamma$ are likely due to {\it NICER}'s narrow energy band, as the photon index is not well  constrained  when the energy spectra are dominated by the disk (soft) component.

Post-outburst reflares (also known as re-brightenings) have been observed in several BHTs, e.g., XTE J1650$-$500 \citep{Tomsick2004}, MAXI J1659$-$152 \citep{Homan2013}, GRS 1739$-$278 \citep{Yan2017}, MAXI J1535$-$571 \citep{Parikh2019,Cuneo2020}, and MAXI J1820+070 \citep{Stiele2020}. The peak fluxes of the reflares are at least $\sim1-2$ orders of magnitude fainter than that of the main outburst. 
For the cases of XTE J1650$-$500, MAXI J1659$-$152 and MAXI J1820+070, the source remained in the HS during the reflares, similar to what we found here in MAXI J1348$-$630.
This type of spectral evolution is reminiscent of the so-called ``failed outbursts" observed in many BHTs \citep[e.g.,][]{Sturner2005,Stiele2016,Capitanio2009,Furst2015,Del2016}.
However, for the case of GRS 1739$-$278, the source reached the SS during the two reflares after its main outburst, and their tracks in the HID show hysteresis \citep{Yan2017}.
Before returning to quiescence, MAXI J1535$-$571 exhibited at least four reflares, and their peak fluxes decayed with time. During the first two brighter reflares, the source transited from the HS into the SS; however, during the third and fourth reflares, MAXI J1535$-$571  did not reach the SS \citep{Parikh2019,Cuneo2020}. 
This result shows that the hard-to-soft transition can take place at a wide range of luminosities. In GX 339$-$4, the hard-to-soft transition luminosities are also different between outbursts, but all transitions occur at luminosity higher than the peak luminosity reached during failed outbursts \citep{Belloni2011}. This suggests that for each source there might be a critical luminosity (linked with a critical accretion rate) for the hard-to-soft transition to occur.


We observed different types of low-frequency QPOs at different phases of the outburst. Type-C QPOs were seen in the HS and HIMS. Type-B and type-A QPOs were detected in the SIMS and SS, respectively. The frequency ranges and $Q$ factors of the QPOs are consistent with that observed in other BHTs with {\it RXTE} \citep[see, e.g.,][]{Motta2016}. 
However, the fractional rms of the QPOs in the \nicer 0.5--12 keV band are lower than the typical values obtained from the {\it RXTE}/PCA 2--13 keV band for PDS with similar shapes. For example, the type-B QPOs we found have a fractional rms of $<1.5\%$ in the \nicer 0.5--12 keV band, while the typical value for the {\it RXTE}/PCA 2--13 keV band is $\sim4-5\%$ \citep[e.g.,][]{Casella2004}.
The 0.5--64 Hz integrated fractional rms we measured in the HS (20--30\%) is also significantly lower than that typically seen in the {\it RXTE}/PCA band (30--40\%, \citealt{Munoz2011}).
Although there is no consensus on the physical origin of the QPOs and broadband noise, it is suggested that their amplitude is set by the physical component that modulates the photons at $\gtrsim2-3$keV \citep[e.g.,][]{Ingram2009,Ingram2012,Stevens2016}. The lower variability we observed with \nicer could be intrinsic to MAXI J1348--630, but most probably is due to dilution of the amplitude caused by the non-modulated photons we observe at $<2$ keV thanks to the effective area of {\it NICER}. 
An 18-Hz QPO was observed at ${\rm hardness}\sim0.01$ accompanied by a relatively strong peaked noise component (Fig. \ref{fig:pds}e). This type of PDS is similar to that found in XTE J1550$-$564 \citep{Homan2001}, H 1743$-$322 \citep{Homan2005b} and GRO J1655$-$40 \citep{Remillard1999,Motta2012}. \citet{Remillard1999} found that these QPOs are stronger at higher energy bands. In the frequency-rms diagram, these QPOs follow the same track as those type-C QPOs seen in the HS and HIMS \citep{Motta2012}.

\section*{Acknowledgements}

We are grateful for the anonymous referee's helpful comments and suggestions. L.Z. acknowledges support from the Royal Society Newton Funds. D.A. acknowledges support from the Royal Society. V.A.C. acknowledges support from the Royal Society International Exchanges ``The first step for High-Energy Astrophysics relations between Argentina and UK" and from the Spanish \textit{Ministerio de Ciencia e Innovaci\'on} under grant AYA2017-83216-P. K.A. acknowledges support from a UGC-UKIERI Phase 3 Thematic Partnership (UGC-UKIERI-2017-18-006; PI: P. Gandhi). F.M.V. acknowledges support from STFC under grant ST/R000638/1. This work was supported by NASA through the \textit{NICER} mission and the Astrophysics Explorers Program. This research has made use of MAXI data provided by RIKEN, JAXA and the MAXI team, and data and software provided by the High Energy Astrophysics Science Archive Research Center (HEASARC) and NASA's Astrophysics Data System Bibliographic Services. 

\section*{Data Availability}

The data underlying this article are available in the HEASARC database.


\appendix
\section{Additional Figure}

\begin{figure}
\begin{center}
\resizebox{1.02\columnwidth}{!}{\rotatebox{0}{\includegraphics{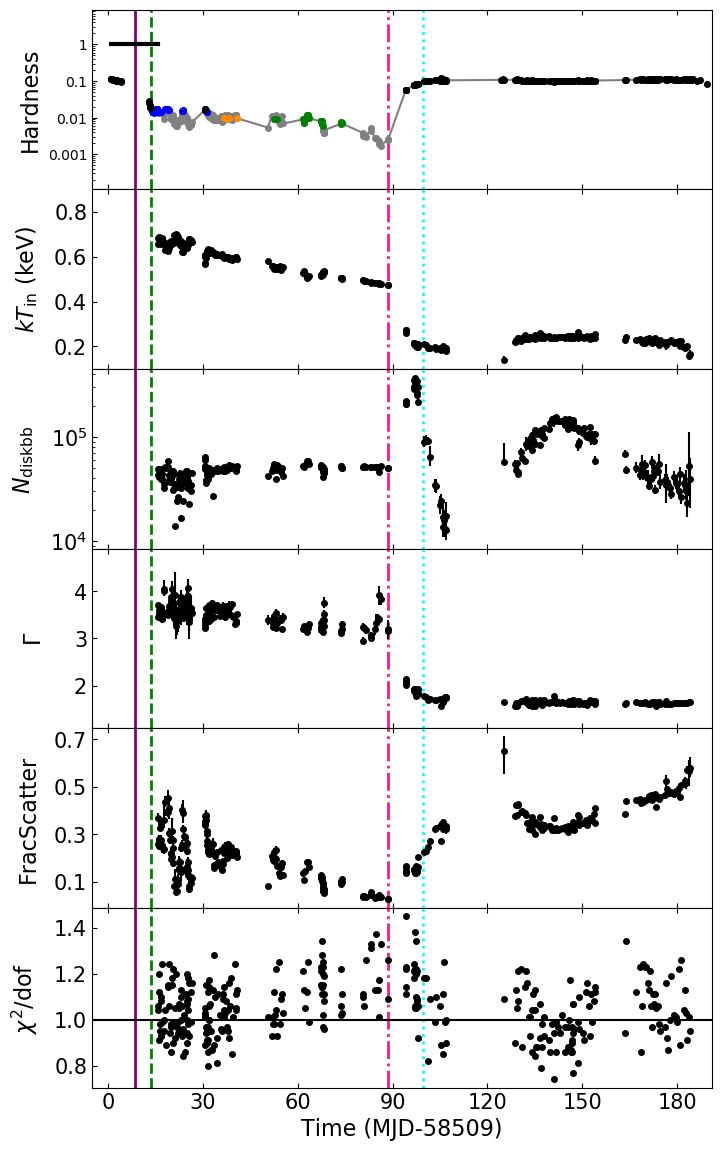}}}
\end{center}
\caption{Evolution of the main spectral parameters from the spectral fits with the continuum model \textsc{tbnew*(simpl*diskbb)}. In the top panel we show the hardness ratio for comparison. From the second panel to the bottom: disk temperature at the inner radius ($kT_{\rm in}$) in keV, \textsc{diskbb} normalisation ($N_{\rm diskbb}$), photon index ($\Gamma$), scattered fraction (FracScatter), and the reduced $\chi^{2}$. Same as Fig. \ref{fig:spectral_para}, we did not plot the data after day 187 (MJD 58596), since that the disk component was not needed in the spectral fits and the spectra can be fitted well with a single power-law with $\Gamma\sim1.6$.}
\label{fig:simpl}
\end{figure}

\bsp	
\label{lastpage}
\end{document}